\documentstyle[lcws2002,psfig,epsfig,endnotes,longtable,pifont,boxedminipage,rotating,axodraw]{article}

\def\be{\begin{equation}}
\def\ee{\end{equation}}
\def\bea{\begin{eqnarray}}
\def\eea{\end{eqnarray}}

\def\v{\begingroup\obeyspaces\u}
\def\u#1{\tt#1\endgroup}
\def\IQ{\v{IQ}}
\def\IL{\v{IL}}

\def\IN{\v{IN}}
\def\IC{\v{IC}}

 \def\ino{\widetilde}
 \def\squark{\ino{q}} 
 \def\slepton{\ino{\ell}}  \def\snu{\ino{\nu}}
 \def\gaugino{\ino{\chi}} 
\def\s#1{{\small#1}}

\def\TA{{\small TAUOLA}}
\def\HW{\s{HERWIG}}

\def\4J{\s{4JPHACT}}
\def\PY{\s{PYTHIA}}
\def\IS{\s{ISAJET}}
\def\IW{\s{ISAWIG}}

\def\MSSM{{MSSM}}
\def\SY{\s{SUSY}}
\def\QCD{\s{QCD}}

\def\ee{e^+e^-}

\def\B0bar{\overline{B^0}}

\def\l{\ell}
\begin{document}

\begin{flushright}
{CERN-TH/2002-237\\
IPPP/02/53\\ 
DCPT/02/106\\
September 2002}
\end{flushright}

\title{HERWIG: AN EVENT GENERATOR \\ FOR $e^+e^-$ LINEAR COLLIDERS}

\author{Stefano Moretti}
\address{CERN Theory Division, CH-1211 Geneva 23, Switzerland
and\\
Institute for Particle Physics Phenomenology, Durham DH1 3LE, UK}

\maketitle\abstracts{
I review all the new features of the HERWIG event
generator which are relevant to Linear Collider (LC)
 physics starting from version 6.1.}

\vspace*{-0.75cm}

\section{Introduction}

    \HW\ is a general-purpose Monte Carlo (MC) event generator for high-energy
    processes \cite{HERWIG},  providing a full simulation of
    hard lepton-lepton, lepton-hadron and  hadron-hadron scattering
    and soft hadron-hadron collisions in a single package, comprising
    the following features.

\noindent
    1. Initial- and final-state \QCD\ jet evolution with soft gluon
    interference taken into account via angular ordering.

\noindent
    2. Colour coherence of (initial and final) partons in all
subprocesses, including the production and decay of 
Supersymmetric (\SY)  particles.

\noindent 
3. Matrix Element (ME) corrections to the parton shower (PS)
algorithm.

\noindent 4. Spin correlations in the decay of heavy fermions.

\noindent 5. Lepton beam polarisation in selected processes.

\noindent
    6.~Azimuthal correlations  within and  between jets due to gluon
interference and polarisation.

\noindent
    7.
A cluster model for jet hadronisation  (with optional colour rearrangement)
based on non-perturbative gluon splitting
and a similar model for soft and underlying
hadronic  events.

\noindent
The \HW\ source codes and related information
can be found in \cite{HWweb}.

\section{Linear Collider physics with HERWIG}

Of particular relevance to the programme of future electron-positron
LCs are items 3.--5. We will describe these in the following, after
mentioning the Minimal Supersymmetric Standard Model
(\MSSM) processes currently available in lepton-antilepton annihilations.

\subsection{\MSSM\ processes in $\ell^+\ell^-$ collisions ($\ell=e,\mu$)}

Starting from version 6.1, 
\HW\ includes the production and decay of (s)particles,
as given by the \MSSM. The conventions used and a
detailed description of the implementation can be found in \cite{SUSYWIG}.

\HW\ does not contain any  built-in models for \SY-breaking  
scenarios. In all cases the general \MSSM\ particle spectrum
and decay tables must be provided just like those for any other
object, with the following caveats:
(i) \SY\ particles do not radiate (which is reasonable if their
decay lifetimes are much shorter than the QCD confinement scale); 
(ii) CP-violating \SY\ phases are not included.
A package, \IW, has been created to work with {\IS} \cite{ISAJET}
to produce a file containing the \SY\ particle masses, lifetimes,
couplings and mixing 
parameters. This package takes  the outputs of the \IS\
\MSSM\ programs and produces a data file in a format that can be read into
\HW\ for the subsequent process generation. The user can
produce her/his own file provided that the correct format is used.
To this end, we invite the consultation of the \IW\ webpage \cite{IWweb},
where some example of input files can be found (the \SY\ benchmark 
points recommended in Ref.~\cite{SPS} are also available). 

\begin{center}
\vskip0.20cm
{\small {\bf Table 1.} 
The \MSSM\ processes via $\ell^+\ell^-$ annihilations implemented in \HW.}
\small
\begin{tabular}{|c|l|}
\hline
 {\tt IPROC} &  \MSSM\ processes \\
\hline
   700-99  & R-parity conserving \SY\ processes \\
   700     & $\l^+ \l^- \to$~2-sparticle processes (sum of 710--760)\\
   710     & $\l^+ \l^- \to$~neutralino pairs (all neutralinos) \\
706+4{\tt IN1}+{\tt IN2} &$\l^+ \l^- \to \gaugino^0_{\mbox{\scriptsize IN1}}
                         \gaugino^0_{\mbox{\scriptsize IN2}}$
                      ({\tt IN1,2}=neutralino mass eigenstate)\\
   730     & $\l^+ \l^- \to$~chargino pairs (all charginos) \\
728+2{\tt IC1}+{\tt IC2} &$\l^+ \l^- \to \gaugino^+_{\mbox{\scriptsize IC1}}
                         \gaugino^-_{\mbox{\scriptsize IC2}}$
                      ({\tt IC1,2}=chargino mass eigenstate) \\
   740     & $\l^+ \l^- \to$~slepton pairs (all flavours) \\
   736+5\IL& $\l^+ \l^- \to \slepton_{L,R} \slepton_{L,R}^*$
             ($\IL=1,2,3$ for $\slepton=\tilde{e},\tilde{\mu},\tilde{\tau}$) \\
   737+5\IL& $\l^+ \l^- \to \slepton_{L} \slepton_{L}^*$ (\IL\ as above) \\
   738+5\IL& $\l^+ \l^- \to \slepton_{L} \slepton_{R}^*$ (\IL\ as above)\\
   739+5\IL& $\l^+ \l^- \to \slepton_{R} \slepton_{R}^*$ (\IL\ as above)\\
   740+5\IL& $\l^+ \l^- \to \snu_{L} \snu_{L}^*$ 
             ($\IL=1,2,3$ for $\snu_e, \snu_\mu, \snu_\tau$) \\
   760      & $\l^+ \l^- \to$~squark pairs (all flavours) \\
   757+4\IQ & $\l^+ \l^- \to \squark_{L,R} \squark^*_{L,R}$
             ($\IQ=1...6$ for $\squark=\tilde{d}...\tilde{t}$)\\
   758+4\IQ & $\l^+ \l^- \to \squark_{L} \squark^*_{L}$
                (\IQ\ as above)\\
   759+4\IQ & $\l^+ \l^- \to \squark_{L} \squark^*_{R}$
                (\IQ\ as above)\\
   760+4\IQ & $\l^+ \l^- \to \squark_{R} \squark^*_{R}$
                (\IQ\ as above)\\
\hline
\end{tabular}
\end{center}

\vskip0.30cm
In addition to the decay modes implemented in the \IS\ package, \IW\ also
includes the calculation of all 2-body squark/slepton 
and 3-body gaugino/gluino R-parity violating (RPV) decay modes 
(alas, RPV lepton-gaugino and slepton-Higgs mixing is not considered).
Moreover,
the emulation of RPV processes is also a feature of the production
stage. Tab.~1 illustrates all \MSSM\ modes initiated by $\ell^+\ell^-$
scattering that are available at present (version 6.4). Here, 
{\tt IPROC} is the input label selecting the hard process.

\begin{center}
{\small {\bf Table~1.} Continues.}\\
\small
\begin{tabular}{|c|l|}
\hline
   800-99  & R-parity violating \SY\ processes \\
   800     & Single sparticle production, sum of 810--840 \\
   810     & $\l^+ \l^- \to \gaugino^0 \nu_i$, (all neutralinos)\\
   810+\IN & $\l^+ \l^- \to \gaugino^0_{\mbox{\scriptsize IN}} \nu_i$,
             (\IN=neutralino mass state)\\
   820     & $\l^+ \l^- \to \gaugino^- e^+_i$ (all charginos) \\
   820+\IC & $\l^+ \l^- \to \gaugino^-_{\mbox{\scriptsize IC}} e^+_i$,
             (\IC=chargino mass state) \\
   830     & $\l^+ \l^- \to \snu_i Z^0$ and 
             $\l^+ \l^- \to \slepton^+_i W^-$  \\
   840     & $\l^+ \l^- \to \snu_i h^0/H^0/A^0$ and 
             $\l^+ \l^- \to \slepton^+_i H^-$  \\
   850     & $\l^+ \l^- \to \snu_i \gamma$ \\
   860     & Sum of 870 and 880 \\
   870     & $\l^+ \l^- \to \l^+ \l^-$, via LLE only \\
   867+3{\tt IL1}+{\tt IL2} & 
$\l^+ \l^- \to \l^+_{\mbox{\scriptsize IL1}} \l^-_{\mbox{\scriptsize IL2}}$
          ({\tt IL1,2}=1,2,3 for $e,\mu,\tau$) \\
   880     & $\l^+ \l^- \to \bar d  d$, via LLE and LQD \\
   877+3{\tt IQ1}+{\tt IQ2} & 
$\l^+ \l^- \to d_{\mbox{\scriptsize IL1}} \bar d_{\mbox{\scriptsize IL2}}$
          ({\tt IQ1,2}=1,2,3 for $d,s,b$) \\
\hline
\end{tabular}
\end{center}

\begin{center}
{\small {\bf Table~1.} Continues.}\\
\small
\begin{tabular}{|c|l|}
\hline
 910-975 &   Higgs processes  \\
    910    &      $\ell^+ \ell^- \to \nu_\ell \bar\nu_\ell h^0 + \ell^+ \ell^- h^0$\\
    920    &      $\ell^+ \ell^- \to \nu_\ell \bar\nu_\ell H^0 + \ell^+ \ell^- H^0$\\
\hline
    960    &      $\ell^+ \ell^- \to Z^0 h^0$\\ 
    970    &      $\ell^+ \ell^- \to Z^0 H^0$\\ 
\hline
    955    &      $\ell^+ \ell^- \to H^+ H^-$\\
    965    &      $\ell^+ \ell^- \to A^0 h^0$\\
    975    &      $\ell^+ \ell^- \to A^0 H^0$\\
\hline
\end{tabular}
\end{center}

\subsection{ME corrections to the standard PS} 

The study of top quark production and decay will be one of the main areas of 
research activity at future LCs. Hence, it is of paramount importance the
availability of a MC event generator describing the dynamics of the
$e^+e^-\to t\bar t\to b\bar b W^+W^-$ process ($m_t=175$ GeV)
with great accuracy. As the top quark is a coloured particle, of particular
concern are higher order effects from QCD. Whereas those induced by
the exchange of virtual gluons do not change the lowest order kinematics,
real radiation of the latter does so. This is of relevance when it comes to
reconstruct the top quark parameters (primarily, $m_t$ and $\Gamma_t$)
from some exclusive jet observables, in presence of $W^\pm\to $ jet-jet
decays. 

\begin{figure}[!h]
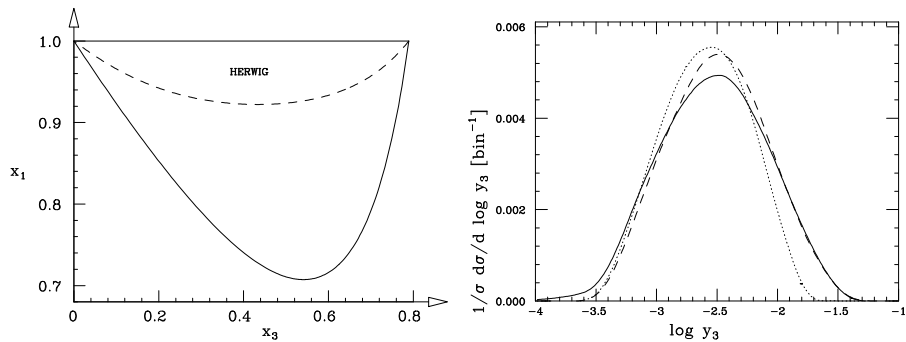

\vskip-0.3cm
\centerline{\resizebox{0.49\textwidth}{!}{\includegraphics{lc0.ps}}%
\hfill%
\resizebox{0.49\textwidth}{!}{\includegraphics{lc1.ps}}}
  \caption{\small The $t\to W_1^+ b_2  g_3$ decay: (Left) The total 
(solid) and 
HERWIG (dashed) 
phase space in terms of the $x_1$ and $x_3$ momentum fractions; 
(Right) The $y_3$ distributions according to the new HERWIG implementation 
(dashed), to the old one (dotted) and to the exact ${\cal O}(\alpha_S)$ 
calculation (solid).
Rates are at $360$~GeV.}
\label{fig:topdecay}
\end{figure}

\begin{figure}[!ht]
\centerline{\resizebox{0.40\textwidth}{!}{\includegraphics{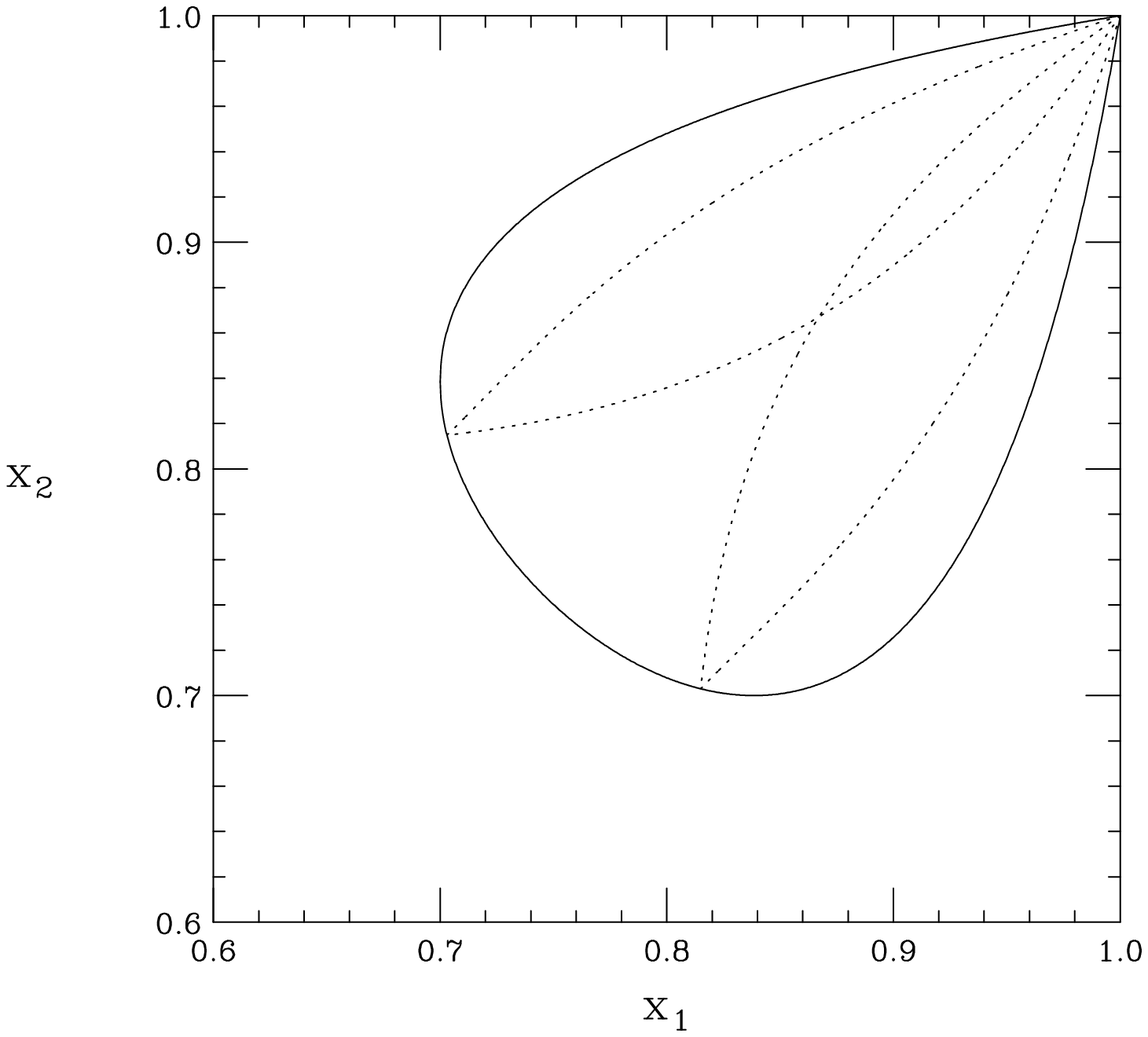}}%
\hfill%
\resizebox{0.49\textwidth}{!}{\includegraphics{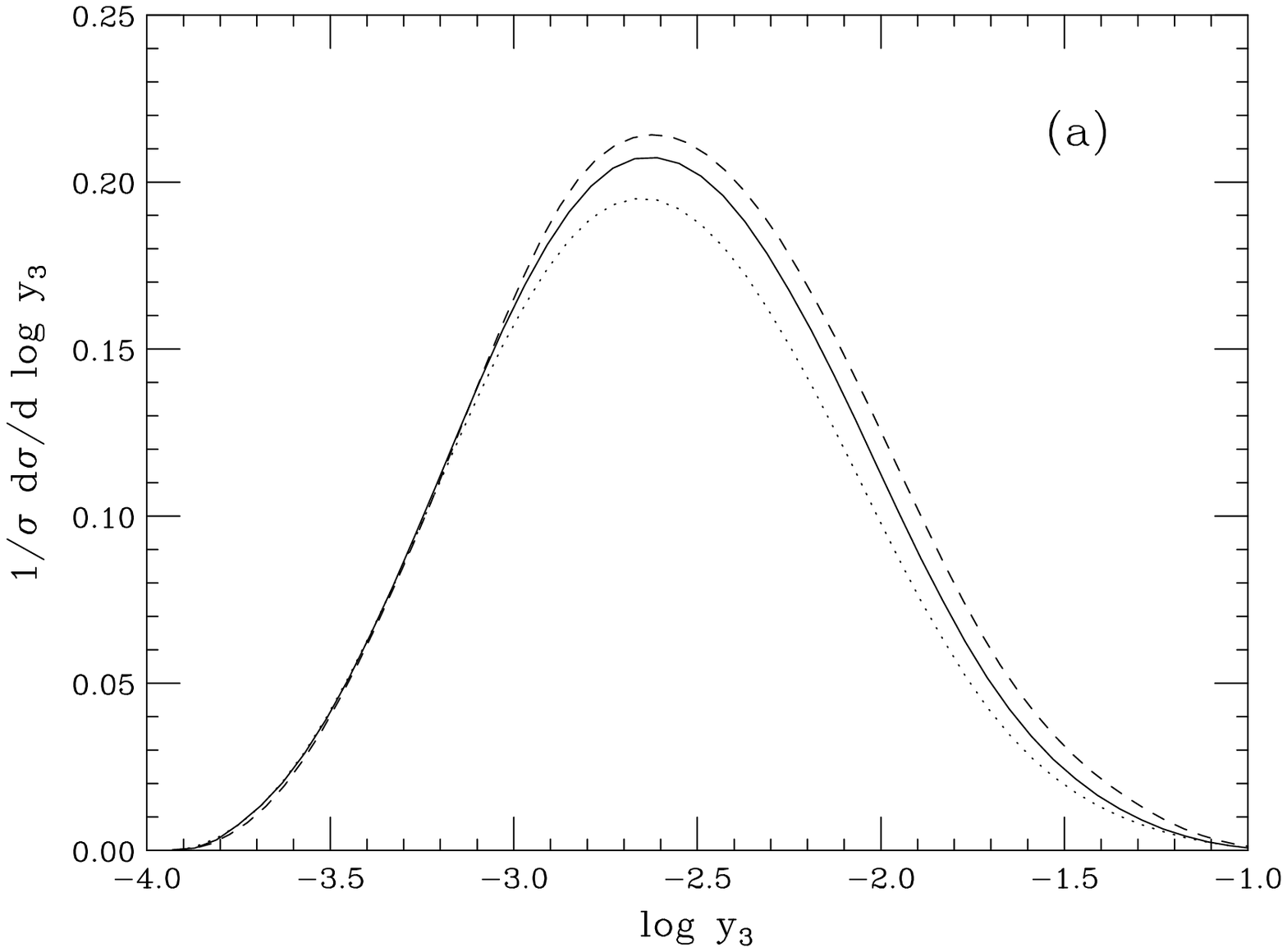}}}
  \caption{\small The $e^+e^-\to t_1 \bar t_2  g_3$ production:
 (Left) 
Total (solid) and HERWIG (dotted) phase space limits for
$t\bar t$ production in terms of the $x_1$ and $x_3$ momentum fractions; 
(Right) The $y_3$ distributions according to the old HERWIG implementation 
(dashed), to the new one, the latter 
populating either the small- and large-angle 
(solid) or only the large-angle region (dotted) of the dead zone.
Rates are at $500$~GeV.}
\label{fig:topproduction}
\vskip-0.3cm
\end{figure}

According to the HERWIG standard algorithm for PS,
gluon radiation is treated in the soft and collinear approximation and 
no emission is permitted in the `dead zones' (hard and 
large-angle parton radiation) seen in the left-hand side of 
Figs.~\ref{fig:topdecay}--\ref{fig:topproduction}. The current version of
the HERWIG event generator has been improved by applying so-called ME
corrections: i.e.,  
the dead zone is populated by the use of the exact first-order matrix element
(`hard correction') and the ${\cal O}(\alpha_S)$ result is used in the 
already-filled region any time an emission is the `hardest so far'
(`soft correction') \cite{MEs}.


A good measure to gauge the improvement due to the new algorithm can be gained
from investigating the threshold value of the Durham jet-clustering algorithm 
\cite{DURHAM} for all events to yield three jets,
each with $E_T>10$~GeV and relative $\Delta R>$~0.7
in transverse energy and cone separation. Just above  
$\sqrt s=2m_t$, where real radiation from the production stage is negligible,
the beneficial effects of the new \HW\ implementation can be appreciated in 
the right-hand side of Fig.~\ref{fig:topdecay}, in the case of top quark 
decays. At large $y_3$ values, where
the fixed order ${\cal O}(\alpha_S)$ description \cite{Tim}
is expected to well describe
the underlying (hard gluon) dynamics, we see the latter to coincide with the
current \HW\ description, which in turns retains sizable differences from
the ${\cal O}(\alpha_S)$  result as
$y_3\to0$ (soft and collinear gluon), where the all-order PS algorithm 
is known to be reliable. While similar improvements can be seen in the
case of top quark production, it is in this instance more instructive to 
compare the same $y_3$ distribution as obtained from the old \HW\ description
(incorporating ME corrections but not the complete $\sim m_t^2/s$ mass effects)
to those given by two more advanced options, currently being implemented: the 
default one filling both the small- and large-angle dead zones
and the intermediate one occupying only the large-angle region, 
both accounting for the above mass corrections
(see the right-hand side of Fig.~\ref{fig:topproduction}). The sizable
differences  between the three clearly makes the point in favour of the more
complete \HW\ implementation in view of future LC studies. 

Another example of ME corrections available in \HW\ is the 
case of multi-jet production in electron-positron annihilations via
exact fixed-order $e^+e^-\to n$~parton MEs, 
limitedly to the case $n=4$. Here, the
event generation can be initiated by the $e^+e^-\to q\bar q gg$
and $e^+e^-\to q\bar q Q\bar Q$ MEs, with the four parton `interfaced'
to the subsequent PS \cite{4jet} (i.e., a hard cut-off is enforced in order
to separate the partons thus preventing the MEs from diverging and the PS is 
attached to each of the latter only over the 
reduced phase space). This is 
an alternative to the standard description based on the 
$e^+e^-\to q\bar q(g)$ 
hard
scattering process followed by the traditional PS algorithm.
While the new implementation is only reliable for $n\ge4$ and
for jet separations larger than the corresponding partonic cut-off, it
remedies the drawback of the old one in describing the angular
orientation of four-jet events, as seen by, e.g., the ALEPH
Collaboration \cite{ALEPH} (see also Ref.~\cite{Oxford}). 

\begin{figure}[!t]
~\hskip2.5cm\epsfig{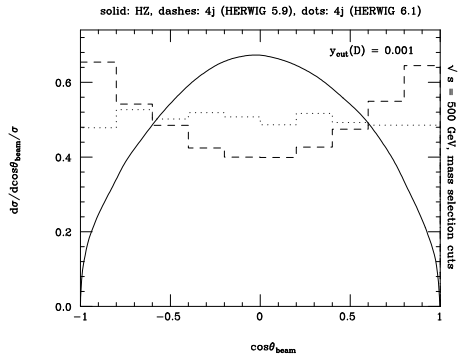}
\caption{\small Distributions for the polar angle
of the reconstructed $Z$ for the Higgs signal and the two description of the 
QCD four-jet background mentioned in the text.}
\label{fig:polar}
\end{figure}

The ability of a MC event generator to describe correctly a four-jet
final state is particularly relevant in the context of LC searches for a
light Higgs state, predominantly produced via $e^+e^-\to ZH\to 4$~jets
events, as BR[$Z\to$ jet-jet]=0.7 and BR[$H\to b\bar b$] $\approx1$. 
Take for example an hadronic sample composed of two $b$-jets and two
untagged jets at a 500 GeV LC, selected by using the 
Durham jet-clustering algorithm with resolution 
$y_{{4}}=0.001$ and with one
pair of jets reproducing the $Z$ mass within 10 GeV, i.e.
$|M_{\mathrm{jj}}-M_Z|<10$ GeV. Then compare the polar angle distribution of 
the di-jet system emulating
a $Z$ decay, for the QCD four-jet background and the above Higgs signal
(for $M_H=150$ GeV): see Fig.~\ref{fig:polar}. It is clear that,
in the region populated by the latter, the 
discrepancy between the old (\HW\ v5.9) and new (\HW\ v6.1) description
of the former is rather large, similar in size to the signal excess. 
Needless to say, under these circumstances,
a clear understanding of the behaviour
of the QCD background is crucial in order to extract the Higgs resonance and 
measure its relevant parameters ($M_H$, $\Gamma_H$, etc.)
\footnote{Options similar to the one available in \HW\ for $n=4$ can also be found 
in {\PY} \cite{PY} and  {\4J} \cite{4JPHACT}.}.

A more sophisticated 
implementation `matching' MEs and PS, meaning the procedure of 
exploiting 
a combined approach which uses MEs for large phase space separations, PS
in the infrared (i.e., soft and collinear) limit and modified (by 
Sudakov form factors) MEs 
in the overlapping region \cite{matching}, is available
in APACIC++, for $n\le6$ \cite{apacic} \footnote{A similar standalone 
implementation for $n\le4$ is available in \HW\ too \cite{Bryan}.}.

\subsection{Spin correlations and lepton beam polarisation}

Since v6.4, spin  correlations  are available in  processes where 
\SY\ particles, top quarks and $\tau$-leptons are produced
\cite{Spin}. Whenever these particles
decay,  2-, 3- and 4-body MEs are used to describe their
dynamics (with or without the spin correlations). In the case 
of $\tau$'s, an interface to {\TA} \cite{TAUOLA} is also available.
Polarisation has been implemented for incoming leptonic beams in
\SY\ processes. These effects are included both
    in the  production  of \SY\  particles  and via  the  spin  correlation
    algorithm in their decays.

As an example of the impact of such spin effects,
consider pair production of the two lightest neutralinos at an (un)polarised
LC, via the following graphs:
\vskip0.25cm
%
%
\begin{figure}[!ht]
\begin{center}
\begin{picture}(300,60)
%
%
\SetOffset(30,0)
\ArrowLine(0,60)(30,30)
\ArrowLine(30,30)(0,0)
\Photon(30,30)(60,30){4}{4}
\ArrowLine(60,30)(90,60)
\ArrowLine(60,30)(90,0)
\Text(-2,60)[r]{${e}^-$}
\Text(-2, 0)[r]{${e}^+$}
\Text(92,60)[l]{$\tilde{\chi}^0_2$}
\Text(92, 0)[l]{$\tilde{\chi}^0_1$}
\SetOffset(170,0)
\ArrowLine(0,60)(45,60)
\ArrowLine(45,60)(90,60)
\DashArrowLine(45,60)(45,0){5}
\ArrowLine(45,0)(0,0)
\ArrowLine(45,0)(90,0)
\Text(-2,60)[r]{${e}^-$}
\Text(-2, 0)[r]{${e}^+$}
\Text(49,30)[l]{$\tilde{{e}}_{L,R}$}
\Text(92,60)[l]{$\tilde{\chi}^0_2$}
\Text(92, 0)[l]{$\tilde{\chi}^0_1$}
\end{picture}
\end{center}
\end{figure}
\vskip0.350cm
\begin{figure}[!ht]
\begin{center}
\begin{picture}(300,60)
\SetOffset(100,0)
\ArrowLine(0,60)(45,60)
\ArrowLine(45,60)(90,60)
\DashArrowLine(45,60)(45,0){5}
\ArrowLine(45,0)(0,0)
\ArrowLine(45,0)(90,0)
\Text(-2,60)[r]{${e}^-$}
\Text(-2, 0)[r]{${e}^+$}
\Text(49,30)[l]{$\tilde{{e}}_{L,R}$}
\Text(92,60)[l]{$\tilde{\chi}^0_1$}
\Text(92, 0)[l]{$\tilde{\chi}^0_2$}
\end{picture}
\end{center}
\end{figure}
\vskip0.15cm
\noindent
with $\tilde\chi_1^0$ the Lightest Supersymmetric Particle (LSP)
and $\tilde\chi_2^0$ decaying via:
\vskip0.25cm
%
%
\begin{figure}[!ht]
\begin{center}
\begin{picture}(300,200)
\SetOffset(20,120)
\SetScale{0.75}
\ArrowLine(-20,60)(40,60)
\ArrowLine(40,60)(80,100)
\Photon(40,60)(80,20){-5}{4.5}
\ArrowLine(80,20)(120,60)
\ArrowLine(120,-20)(80,20)
\Text(-18,46)[r]{$\tilde{\chi}^0_2$}
\Text(62,77)[l]{$\tilde{\chi}^0_1$}
\Text(93,46)[l]{${f}$}
\Text(93,-15)[l]{$\bar{{f}}$}
\Text(40,25)[r]{${Z}^0$}
\SetOffset(20,0)
\SetScale{0.75}
\ArrowLine(-20,60)(40,60)
\ArrowLine(40,60)(80,100)
\DashArrowLine(80,20)(40,60){4.5}
\ArrowLine(80,20)(120,60)
\ArrowLine(120,-20)(80,20)
\Text(-18,46)[r]{$\tilde{\chi}^0_2$}
\Text(62,77)[l]{${f}$}
\Text(93,46)[l]{$\tilde{\chi}^0_1$}
\Text(93,-15)[l]{$\bar{{f}}$}
\Text(45,25)[r]{${\tilde{f}}_{L,R}$}
\SetOffset(200,120)
\SetScale{0.75}
\ArrowLine(-20,60)(40,60)
\ArrowLine(40,60)(80,100)
\DashLine(40,60)(80,20){4.5}
\ArrowLine(80,20)(120,60)
\ArrowLine(120,-20)(80,20)
\Text(-18,46)[r]{$\tilde{\chi}^0_2$}
\Text(62,77)[l]{$\tilde{\chi}^0_1$}
\Text(93,46)[l]{${f}$}
\Text(93,-15)[l]{$\bar{{f}}$}
\Text(40,25)[r]{${h}^0,{H}^0,{A}^0$}
\Text(-30,-32.5)[r]{($f = e,\mu$)}
\SetOffset(200,0)
\SetScale{0.75}
\ArrowLine(-20,60)(40,60)
\ArrowLine(80,100)(40,60)
\DashArrowLine(40,60)(80,20){4.5}
\ArrowLine(80,20)(120,60)
\ArrowLine(80,20)(120,-20)
\Text(-18,46)[r]{$\tilde{\chi}^0_2$}
\Text(62,77)[l]{$\bar{{f}}$}
\Text(93,46)[l]{${f}$}
\Text(93,-15)[l]{$\tilde{\chi}^0_1$}
\Text(45,25)[r]{${\tilde{f}}_{L,R}$}
\end{picture}
\end{center}
\end{figure}
\vskip0.5cm

\begin{figure}
\begin{center}  
{\includegraphics[width=0.45\textwidth,angle=90]{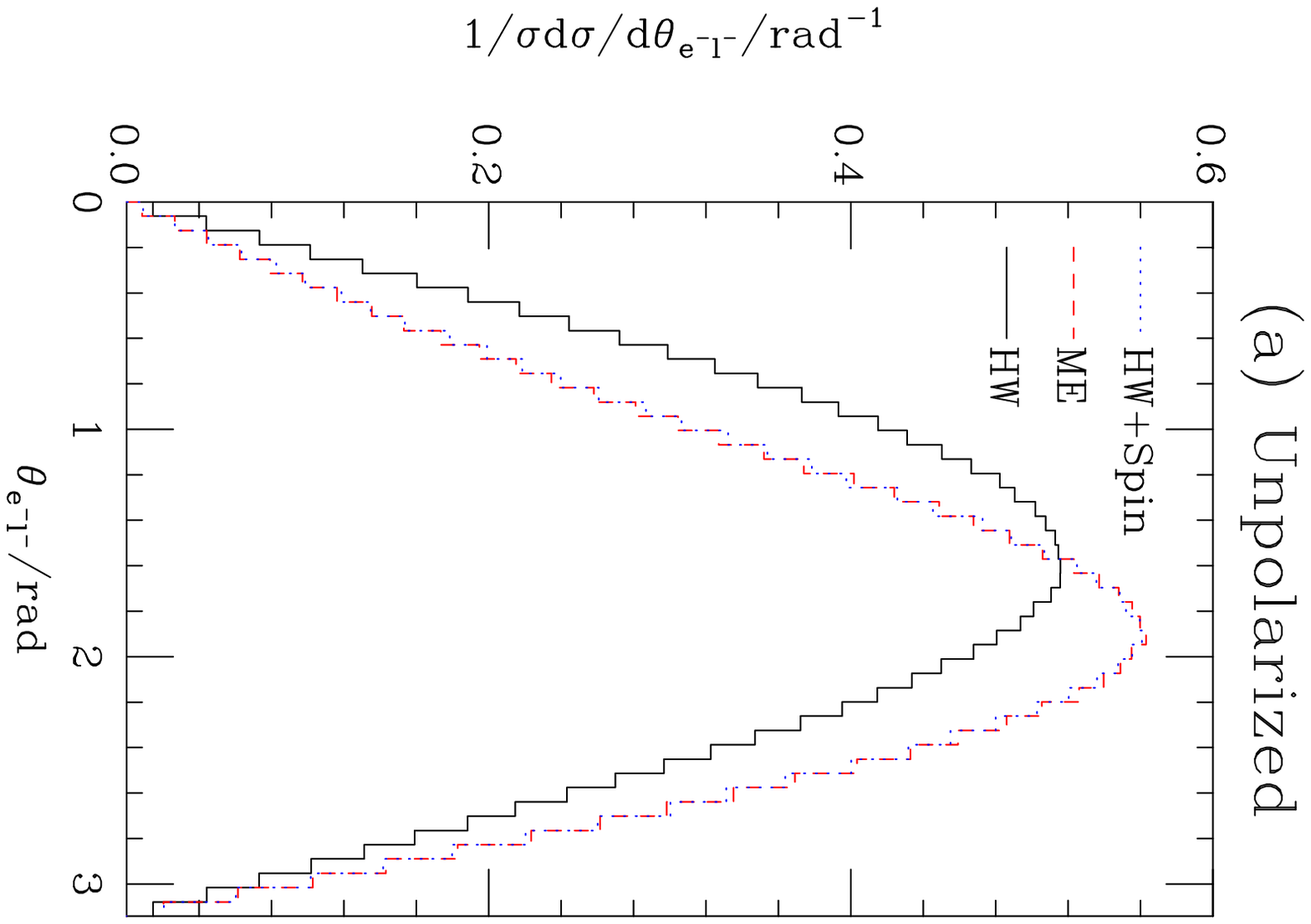}\hfill
 \includegraphics[width=0.45\textwidth,angle=90]{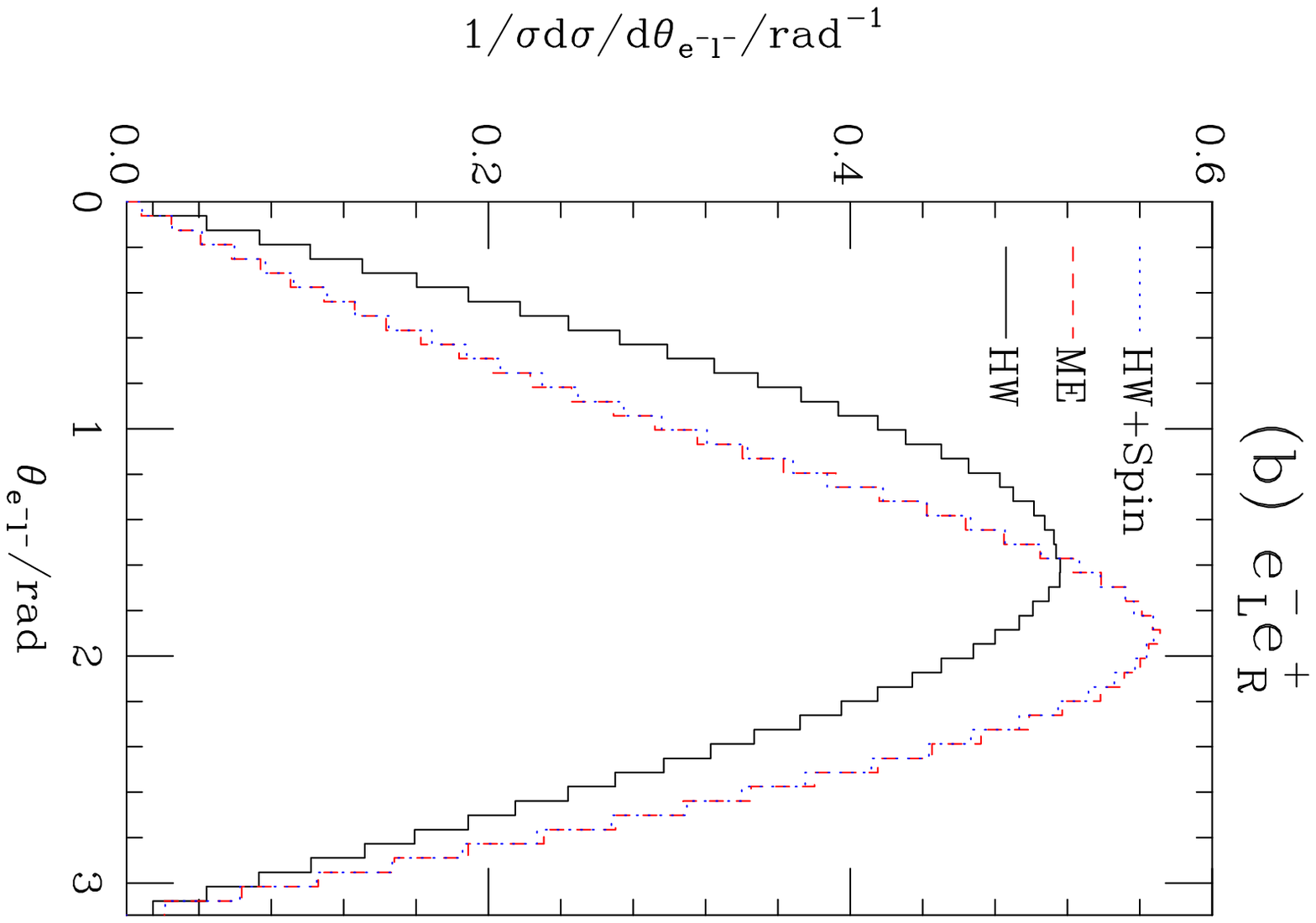}\hfill
 \includegraphics[width=0.45\textwidth,angle=90]{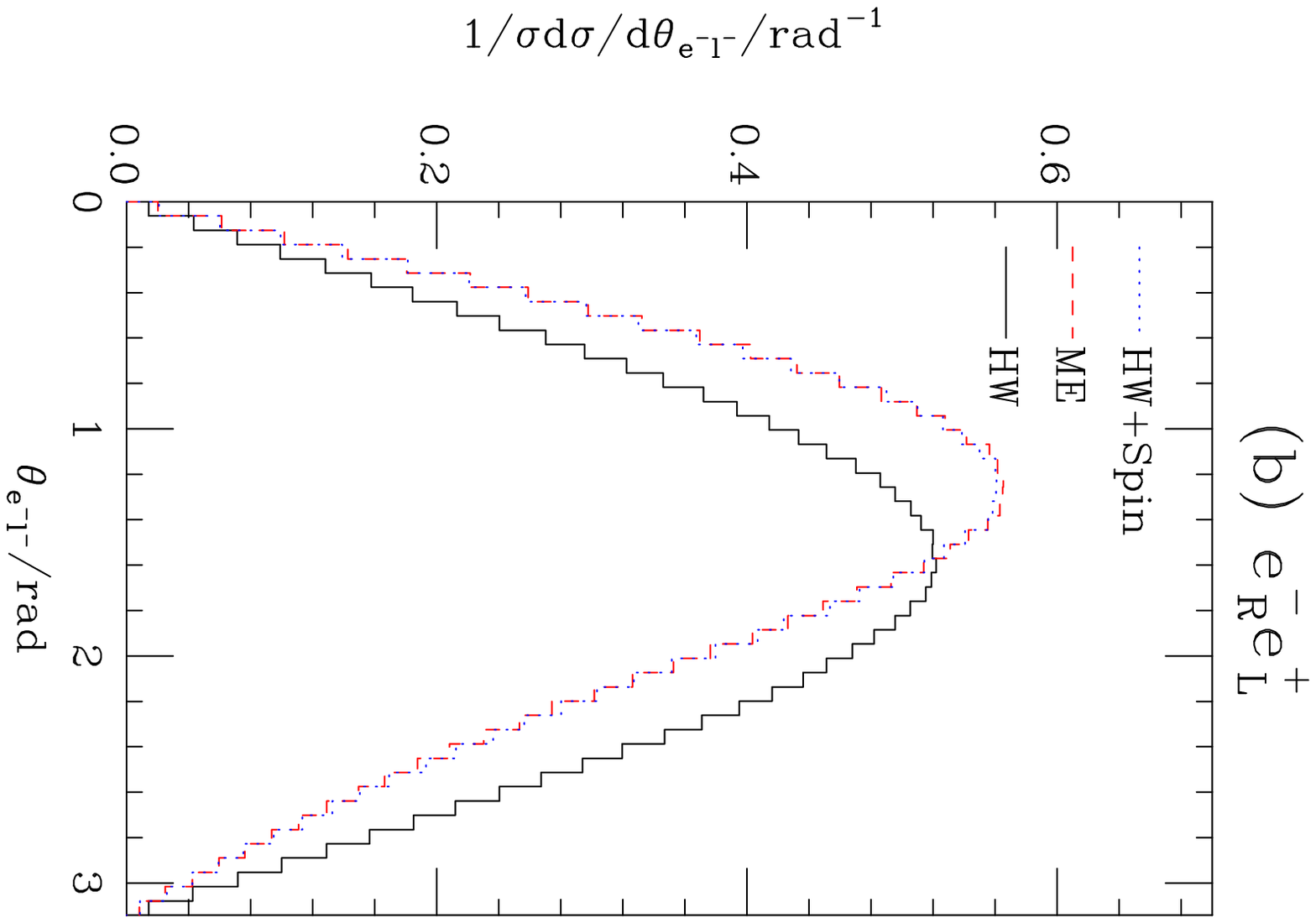}}
\caption{\small Angle between the lepton produced in 
	${e}^+{e}^-\to\tilde{\chi}^0_2
\tilde{\chi}^0_1\to {f}\bar f\tilde{\chi}^0_1\tilde{\chi}^0_1$
($f=e,\mu$)	 and the incoming electron beam in the
  	laboratory frame for $\sqrt s=500$ GeV with:
	{(a)} no polarization, {(b)} negatively polarized electrons and positively polarized positrons 
	and {(c)} positively polarized electrons and negatively polarized positrons.
        The solid line shows the default result from HERWIG which 
	treats the production and decay as independent,
	the dashed line gives the full result from the four-body
	ME and the dotted line the result of the spin correlation algorithm.
(The \SY\ parameters are: $M_0=210$ GeV,
$A_0=0$, $\tan\beta=10$, $M_1=450$ GeV, $M_{2,3}=350$ GeV.)}
\label{fig:spin}
\end{center}
\end{figure}

From Ref.~\cite{Spin}, we show Fig.~\ref{fig:spin},
illustrating the angular behaviour of the above \SY\ signals, 
as given by a typical MC implementation ({\tt HW}), which does
not include spin effects in production and decay and/or ME corrections in 
the latter, to the output of \HW\ v6.4 ({\tt HW+Spin}).  Once again, the improvement is clear
if one confronts the latter with the distributions obtained from a complete
calculation ({\tt ME}), wherein the event generation is based on the {exact} $2\to 6$ ME 
description. 

\vskip0.25cm
I thank the LCWS2002 organisers 
for the excellent atmosphere and stimulating environment that they 
have created during the workshop and The Royal Society of London, UK,
for partial financial support in the form of a Conference Grant.


\begin{thebibliography}{99}

\bibitem{HERWIG}
G. Marchesini, B.R. Webber,  G. Abbiendi, I.G. Knowles, M.H. Seymour
and L. Stanco, {\it Comput.\ Phys.\ Commun.} {\bf 67},  465 (1992);
G. Corcella, I.G. Knowles, G. Marchesini,  
S. Moretti, K. Odagiri, P. Richardson, M.H. Seymour and  B.R. Webber,
 {\tt hep-ph/9912396}; {\it JHEP} {\bf 01},  010 (2001);
 {\tt hep-ph/0107071}; {\tt hep-ph/0201201}.

\bibitem{HWweb}
See: {\tt http://hepwww.rl.ac.uk/theory/seymour/herwig/.}

\bibitem{SUSYWIG} S.\ Moretti, K.\ Odagiri, P.\ Richardson, 
M.H.\ Seymour and B.R.\ Webber, {\it JHEP} {\bf 04},  028 (2002); 
S. Moretti, {\tt hep-ph/0205105}. 
                                                   
\bibitem{ISAJET}
F.E. Paige, S.D. Protopopescu, H. Baer and X. Tata,
{\tt hep-ph/9804321}; {\tt hep-ph/} {\tt 9810440}.

\bibitem{IWweb}
See: \vspace*{-0.70cm} 
\begin{verbatim}
    http://www.hep.phy.cam.ac.uk/~richardn/HERWIG/ISAWIG/.
\end{verbatim}

\vspace*{-0.25cm} 
\bibitem{SPS}
B.C. Allanach {\it et al.},  {\it Eur. Phys. J.} C {\bf 25}, 113 (2002).

\bibitem{MEs} G. Corcella and M.H. Seymour, 
{\it Phys. Lett.} B {\bf 442},  417 (1998);
{\tt hep-ph} {\tt /9911335}; G. Corcella,
E.K. Irish and M.H. Seymour,  {\tt hep-ph/0012319}.

\bibitem{DURHAM} Yu.L.\ Dokshitzer, contribution cited in the `Report of the 
Hard QCD Working Group', in
Proceedings of the Workshop `Jet Studies at LEP and HERA',
       Durham, December 1990, {\it J. Phys} {G {\bf 17}},  1537 (1991);\\
S.~Catani, Yu.L.~Dokshitzer, M.~Olsson, G.~Turnock and B.R.~Webber,
{\it Phys. Lett.} {B {\bf 269}},  {432} ({1991}).

\bibitem{Tim} L.H. Orr, T. Stelzer and W.J. Stirling, 
{\it Phys. Lett.} B {\bf 354}, 442 (1995); {\it Phys. Rev.} D
{\bf 56},  446 (1997).

\bibitem{4jet} S. Moretti, {\tt hep-ph/000819}.

\bibitem{ALEPH} ALEPH Collaboration, {\it Z. Phys.} C {\bf 76},  1 (1997).

\bibitem{Oxford} A. Ballestrero {\it et al.}, {\it J. Phys.} G
{\bf 24},  365 (1998);
S. Moretti and W.J. Stirling, {\it Eur. Phys. J.} C {\bf 9},  81 (1999).

\bibitem{PY} T. Sj\"ostrand, L. L\"onnblad and S. Mrenna,
{\tt  hep-ph/0108264}.

\bibitem{4JPHACT} A. Ballestrero, in preparation.

\bibitem{matching} S. Catani, F. Krauss, R. Kuhn and B.R. Webber,
{\it JHEP} {\bf 11},  063 (2001).

\bibitem{apacic} F. Krauss, R. Kuhn and G. Soff,
{\it J. Phys.} {G {\bf 26}},  L11 (2000); {\it Acta Phys. Polon.}
{B {\bf 30}},  3875 (1999).

\bibitem{Bryan} See: {\tt http://webber.home.cern.ch/webber/.}

\bibitem{Spin}
P.~Richardson, {\it JHEP} {\bf 11},  029 (2001).

\bibitem{TAUOLA}
S. Jadach, Z. Was, R. Decker and J.H. Kuhn,
Comput. {\it Phys. Commun.} {\bf 76},  361 (1993); 
M. Jezabek, Z. Was, S. Jadach and J.H. Kuhn,
Comput. {\it Phys. Commun.} {\bf 70},  69 (1992); 
S. Jadach, J.H. Kuhn and Z. Was,
Comput. {\it Phys. Commun.} {\bf 64},  275 (1990). 



\end{thebibliography}
\end{document}